\documentclass[hyper]{JHEP} 

\usepackage{epsfig}

\newcommand\fverb{\setbox\pippobox=\hbox\bgroup\verb}

\newcommand\fverbdo{\egroup\medskip\noindent%

            \fbox{\unhbox\pippobox}\ }

\newcommand\fverbit{\egroup\item[\fbox{\unhbox\pippobox}]}

\newbox\pippobox

\title{ Hamiltonian Analysis of $1+1$ dimensional  Massive Gravity}
\author{J. Kluso\v{n}\\
Department of
Theoretical Physics and Astrophysics\\
Faculty of Science, Masaryk University\\
Kotl\'{a}\v{r}sk\'{a} 2, 611 37, Brno\\
Czech Republic\\
E-mail: \email{klu@physics.muni.cz}}
\preprint{}

 \abstract{We perform the Hamiltonian
 analysis of $1+1$ dimensional  non-linear
 massive gravity studied  in
arXiv:107.3820. We find the constraint
structure of given theory and perform
the counting of the physical degrees of
freedom. } 

\def\be{\begin{equation}}

\def\ee{\end{equation}}

\def\bea{\begin{eqnarray}}

\def\eea{\end{eqnarray}}

\def\tmH{\tilde{\mH}}

\def\mH{\mathcal{H}}

\def\partt{\partial_\tau}
\def\parts{\partial_\sigma}

\newcommand{\bT}{\mathbf{T}}

\newcommand{\mL}{\mathcal{L}}

\def\pb #1{\left\{#1\right\}}

\begin{document}
\section{Introduction and Summary}\label{first}
One of the most challenging  problems
of modern theory of gravity  is to
decide whether it is possible to find
consistent formulation of massive
gravity.
 The first attempt for
construction of  this theory is dated
to the year 1939 when Fierz and Pauli
formulated its version of linear
massive gravity \cite{Fierz:1939ix}
\footnote{For review, see
\cite{Hinterbichler:2011tt}.}.
The central problem of theories of
massive gravity is that they suffer
from the problem of the ghost
instability, for very nice review, see
\cite{Rubakov:2008nh}. Since the
general relativity is completely
constrained system there are four
constraint equations along the four
general coordinate transformations that
enable to eliminate four of the six
propagating modes of the metric, where
the propagating mode corresponds to a
pair of conjugate variables.  As a
result the number of physical degrees
of freedom is equal to two which
corresponds to the massless graviton
degrees of freedom. On the other hand
in case of the massive gravity the
diffeomorphism invariance is lost and
hence the theory contains six
propagating degrees of freedom which
only five correspond to the physical
polarizations of the massive graviton
while the additional mode is ghost.


De Rham and Gobadadze argued recently
in \cite{deRham:2010ik} that it is
possible to find such a formulation of
the massive gravity which is ghost free
in the decoupling limit. Then  it was
shown in \cite{deRham:2010kj} that this
action that was written in the
perturbative form can be resumed into
fully non-linear actions. It was
claimed there that this is the first
successful construction of potentially
ghost free non-linear actions of
massive gravity.

However it is still open problem
whether this theory contains ghost or
not, see for example
\cite{Chamseddine:2011mu}. On the other
hand S.F. Hassan and R.A. Rosen argued
recently in series of papers
\cite{Hassan:2011vm,Hassan:2011hr,Hassan:2011tf,Hassan:2011zd}
 on the
non-perturbative level that it is
possible to perform such a redefinition
of the shift function so that the
resulting theory still contains the
Hamiltonian constraint. Then it was
argued that the presence of this
constraint allows to eliminate the
scalar mode and hence the resulting
theory is the ghost free massive
gravity. This result was however
questioned in \cite{Kluson:2011qe}
where it was shown that the Hamiltonian
constraint is the second class
constraint which however implies
 that all
non physical modes cannot be
eliminated.

On the other hand it was  argued
recently  in \cite{deRham:2011rn} that
there exists formulation of the ghost
free  massive gravity using its
St\"{u}ckelberg formulation.  This
version of massive gravity  certainly
deserves very careful Hamiltonian
analysis which is however very
difficult due to the complexity of
given action. On the other hand simpler
toy model of $1+1$ dimensional massive
gravity was presented in
\cite{deRham:2011rn} which could  be
explicitly analyzed. In fact, some
preliminary analysis of this model was
given in \cite{deRham:2011rn} but we
feel that it deserves more careful
treatment. Explicitly we mean that  it
is necessary to carefully  identify the
collections of the primary constraints
where we have to take into account the
fact that the $1+1$ gravity is non
dynamical. This is an important
difference from the analysis given in
\cite{deRham:2011rn} where this fact
was not considered. Then we proceed by
the standard way and we find collection
of the secondary constraints. We
analyze   whether these constraints are
preserved during the time evolution of
the system and we find that no new
additional constraints are generated.
This result is again different  from
the conclusion presented in
\cite{deRham:2011rn}. Now using this
known structure of constrains we
determine the number of physical
degrees of freedom. Remarkably we find
that there are no physical degrees of
freedom left exactly as in the standard
case of  scalar theory coupled to two
dimensional gravity. In other words we
confirm the results obtained  in
\cite{deRham:2011rn} even if we derive
them in different  way.

We would like to  stress that $1+1$
dimensional massive gravity is rather
special due to the fact that $1+1$
dimensional gravity is non-dynamical.
For that reason we mean that the result
that all non physical degrees of
freedom are eliminated  should be taken
with care. We expect  that the
situation will be different in case of
higher dimensional massive gravities
due to the fact that now the gravity
becomes dynamical. It would be
certainly nice to see whether the
massive gravity in St\"{u}ckelberg
picture is ghost free or not especially
in the context of the  recent results
found in \cite{Kluson:2011qe}. We hope
to return to the problem of the
Hamiltonian formulation of four
dimensional  the massive gravity in
St\"{u}ckelberg picture in future.

The structure of this note is as
follows. In the next section we perform
the Hamiltonian analysis of the toy
model of the $1+1$ dimensional massive
gravity introduced in
\cite{deRham:2011rn}. In Appendix
(\ref{app}) we present the Hamiltonian
analysis of the model that consists
from two scalar fields minimally
coupled to $1+1$ dimensional  gravity
in order to see the difference with the
Hamiltonian analysis  of $1+1$
dimensional  non-linear massive gravity
 studied in
the main body of the paper.

\section{ $1+1$ dimensional Massive Gravity
 in St\"{u}ckelberg Picture}
Four dimensional non-linear massive
gravity action that was introduced in
\cite{deRham:2010ik} consists from two
parts. The first one is the  standard
Einstein-Hilbert action while the
second one is a specific form of the
Lagrangian for the scalar fields
$\phi^A$ where $A=0,1,2,3$, for more detailed
treatment, see 
\cite{deRham:2010ik}.
 While it is
very difficult to find the Hamiltonian
formulation  of the four dimensional
non-linear massive gravity with
St\"{u}ckelberg fields it is much
easier task to perform the Hamiltonian
analysis of the toy model of non-linear
massive gravity action that was
introduced in \cite{deRham:2011rn}.
Before we proceed to the explicit
analysis of this theory we introduce
following  notation where
$\gamma_{\alpha\beta}$ is a
two-dimensional  metric and $
\sigma^\alpha \ , \alpha,\beta=0,1 \ ,
\sigma^0=\tau \ , \sigma^1=\sigma$ are
corresponding coordinates. Then in
order to formulate the Hamiltonian
formalism  it is convenient to use
ADM formalism
\cite{Arnowitt:1962hi}
\footnote{For review of ADM formalism, see 
\cite{Gourgoulhon:2007ue}.}
 for
 the $1+1$ dimensional
metric
\begin{equation}
\gamma_{\alpha\beta}= \left(
\begin{array}{cc}
-n^2_\tau+ \frac{1}{\omega }n_\sigma^2
&
n_\sigma \\
 n_\sigma & \omega \\
\end{array}\right) \ ,
\end{equation}
where $n_\tau$ is the lapse,
$n_\sigma$ is the  shift and
$\omega$ is spatial part of
metric. Then it is easy to see that
\begin{equation}
\det \gamma=-n_\tau^2 \omega \ , \quad
\gamma^{\alpha\beta}=
\left(\begin{array}{cc}
-\frac{1}{n^2_\tau}
 & \frac{n^\sigma}{ n^2_\tau} \\
 \frac{n^\sigma}{ n^2_\tau}
 &
 \frac{1}{\omega}-
 \frac{n^\sigma n^\sigma}{n_\tau^2  }
 \\ \end{array}\right) \ ,
\end{equation}
where we defined
\begin{equation}
n^\sigma\equiv \frac{n_\sigma}{\omega}
\ .
\end{equation}
%
The non-linear massive gravity action
arises from the coupling of the $1+1$
dimensional gravity to two scalar
fields $\phi^a,a=0,1$ where the scalar
field Lagrangian density has the form
\begin{eqnarray}\label{Lmatter}
\mL_m&=&2n_\tau\sqrt{\omega} [-1+
\sqrt{(D_-\phi^-)(D_+ \phi^+)}=
\nonumber \\
&=&2n_\tau
\sqrt{\omega}\left[-1+\frac{1}{2\lambda}
(D_-\phi^-)(D_+\phi^+)+\frac{\lambda}{2}\right]
\ ,
\nonumber \\
\end{eqnarray}
where
\begin{eqnarray}
\phi^\pm=\phi^0\pm \phi^1 \ , \quad
D_\pm=\frac{1}{\sqrt{\omega}}
\partial_\sigma \pm \frac{1}{n_\tau}
[\partial_\tau-n^\sigma\partial_\sigma
] \ .
\nonumber \\
\end{eqnarray}
In order to see the equivalence of
these two  forms of the Lagrangian
densities given above note that  the
equation of motion for $\lambda $ gives
\begin{equation}
\lambda^2=(D_+\phi^+)(D_-\phi^-) \ .
\end{equation}
Inserting this result into the
expression on the second line in
(\ref{Lmatter}) we recover the form of
the Lagrangian given on the first line.

Now we are ready to proceed to the
Hamiltonian analysis of the action
$S=\int d\tau d\sigma \mL_{m}$, where
$\mL_{m}$ is  given in (\ref{Lmatter}).
As the first step we introduce the
momenta
$\pi^\tau,\pi_\sigma,\pi^\omega,p_\pm$
conjugate to $n_\tau,n^\sigma,\omega$
and $\phi^\pm$ that have non-zero
Poisson brackets
\begin{eqnarray}
\pb{n_\tau(\sigma),\pi^\tau(\sigma')}&=&
\delta(\sigma-\sigma') \ , \quad
\pb{n^\sigma(\sigma),\pi_\sigma(\sigma')}=
\delta(\sigma-\sigma') \ , \nonumber \\
 \pb{\omega(\sigma),
\pi^\omega(\sigma')}&=&\delta(\sigma-
\sigma') \ , \quad
\pb{\lambda(\sigma),\pi_\lambda(\sigma')}=
\delta(\sigma-\sigma')
 \ , \nonumber \\
\pb{\phi^+(\sigma),p_+(\sigma')}&=&\delta(\sigma
-\sigma') \ , \quad \pb{\phi^-(\sigma),
p_-(\sigma')}=\delta(\sigma-\sigma')
 \quad \ .
 \nonumber \\
\end{eqnarray}
Now due to the fact that the $1+1$
dimensional gravity is non-dynamical we
find following primary constraints
\begin{equation}\label{pricon}
\pi^\tau=\frac{\delta S}{\delta \partt
n_\tau}\approx 0 \ , \quad
\pi_\sigma=\frac{\delta S}{\delta
\partt n^\sigma}\approx 0 \ , \quad
\pi^\omega=\frac{\delta S}{\delta\partt
  \omega}\approx 0 \ .
\end{equation}
Finally, from (\ref{Lmatter}) we find
the momenta $p_\pm,\pi_\lambda$
conjugate to $\phi^\pm$ and $\lambda$
\begin{eqnarray}
p_\lambda\approx 0 \ , \quad
p^+=\frac{\delta L}{\delta\partt
\phi^+}=\sqrt{\omega}\frac{1}{\lambda}D_-\phi^-
\ , \quad  p^-=\frac{\delta L}{\delta\partt
\phi^-}=-\sqrt{\omega}\frac{1}{\lambda}D_-\phi^-
\ .
 \nonumber \\
\end{eqnarray}
Using these results we easily derive
the Hamiltonian density
\begin{equation}
\mH=n_\tau \mH_\tau+ n^\sigma
\mH_\sigma-2+ \Gamma_\tau
\pi^\tau+\Gamma^\sigma
\pi_\sigma+\Gamma_\omega\pi^\omega+\Gamma_\lambda
\pi^\lambda\ ,
\end{equation}
where
\begin{eqnarray}
\mH_\tau&=&
\frac{1}{\sqrt{\omega}}(p_-\parts\phi^--
p_+\parts \phi^+)+ 2\sqrt{\omega}
-\lambda\left[\frac{1}{\sqrt{\omega}}p^+p^-+\frac{\sqrt{\omega}}{2}\right]
\ , \nonumber \\
\mH_\sigma&=&
-2\omega\nabla_\sigma\pi_\omega+
p_+\parts \phi^++p_-\parts \phi^-
 \
,\nonumber \\
\end{eqnarray}
and where
$\Gamma_\tau,\Gamma^\sigma,\Gamma_\omega,\Gamma_\lambda$
are Lagrange multipliers corresponding
to the primary constraints
\begin{equation}\label{primcon}
\pi^\tau\approx 0 \ , \quad
\pi_\sigma\approx 0 \ , \quad
\pi^\omega\approx 0 \ , \quad
\pi^\lambda\approx 0 \ .
\end{equation}
Note that  we used the freedom in the
form of the Lagrange multiplier when we
added the primary constraint
$\pi^\omega$ into the definition of the
Hamiltonian. Explicitly,  we added
following expression to the Hamiltonian
$\Gamma'_\omega\pi^\omega=-2n^\sigma
\omega
\nabla_\sigma\pi^\omega+\Gamma_\omega
\pi^\omega$, where $\nabla_\sigma$ is
one dimensional spatial covariant
derivative.  The reason for such a form
of the Lagrange multiplier will be
clear from the  analysis of the time
evolution of the  secondary
constraints. More precisely,
 the requirement of the preservation
of the primary constraints imply secondary
ones
\begin{eqnarray}
\partt \pi^\tau&=&\pb{\pi^\tau,H}=
-\mH_\tau\approx 0 \ , \nonumber \\
\partt \pi^\sigma &=&\pb{\pi^\sigma,H}=
-\mH_\sigma \approx 0 \ , \nonumber \\
\partt \pi^\lambda&=&\pb{\pi^\lambda,H}=
-\frac{1}{\sqrt{\omega}}p^+p^--\frac{1}{2}\sqrt{\omega}=
-G_\lambda \approx 0 \ , \nonumber \\
\partt\pi^\omega&=&\pb{\pi^\omega,H}
\approx
\frac{n_\tau}{2\omega}\lambda\equiv 0 \
.
\nonumber \\
\end{eqnarray}
Since we presume that two dimensional
metric is non-singular we have that
$\omega\neq 0, n_\tau\neq 0 $ and
consequently  the last equation implies
the secondary constraint
$\lambda\approx 0$.  As a result we
have following collection of secondary
constraints $\tmH_\tau\approx 0 \ ,
\mH_\sigma\approx 0 \ ,
G_\lambda\approx 0 \ , \lambda\approx
0$, where  we defined an independent
constraint $\tmH_\tau$ as
\begin{equation}
\tmH_\tau=
\frac{1}{\sqrt{\omega}}(p_-\parts\phi^--
p_+\parts \phi^+)+ 2\sqrt{\omega} \ .
\end{equation}
As the next step we introduce the
smeared form of the constraints
$\tmH_\tau,\mH_\sigma$
\begin{equation}
\bT_\sigma(N^\sigma)=\int d\sigma
N^\sigma\mH_\sigma \ , \quad
\bT_\tau(N)=\int d\sigma
 N\tmH_\tau \ .
 \end{equation}
Note that $\bT_\sigma(N^\sigma)$ has
these  non-zero Poisson brackets
with canonical
variables\begin{eqnarray}
\pb{\bT_\sigma(N^\sigma),\omega}&=&
-\parts \omega N^\sigma -2\omega \parts
N^\sigma \ ,
\nonumber \\
\pb{\bT_\sigma(N^\sigma),p_\pm}&=&
-\parts (N^\sigma  p_\pm) \ , \nonumber
\\
\pb{\bT_\sigma(N^\sigma),\phi^\pm}&=&
-N^\sigma\parts \phi^\pm \ . \nonumber
\\
\end{eqnarray}
Then it is easy to determine following
Poisson brackets
\begin{eqnarray}
\pb{\bT_\sigma(N^\sigma),\bT_\tau(M)}&=&
\bT_\tau(N^\sigma\parts M) \ ,
\nonumber \\
\pb{\bT_\sigma(N^\sigma)
,\bT_\sigma(M^\sigma)}&=& \bT_\sigma
(N^\sigma\parts M^\sigma-M^\sigma\parts
N^\sigma) \nonumber \\
\pb{\bT_\tau(N),\bT_\tau(M)}&=&
\bT_\sigma\left(\frac{1}{\omega}(N\parts
M-M\parts N)\right) \ .
\nonumber \\
\end{eqnarray}
In other  words
$\bT_\sigma(N^\sigma),\bT_\tau(N)$ form
the closed algebra of constraints.
Finally we list non-zero  Poisson
brackets between the constraint
$G_\lambda$ and remaining  constraints
\begin{eqnarray}
\pb{\bT_\sigma(N^\sigma),G_\lambda}&=&
-\parts G_\lambda N^\sigma-\parts
N^\sigma G_\lambda \ , \nonumber
\\
\pb{\bT_\tau(N),G_\lambda}&=&
\frac{N}{\omega} (\parts p^- p^+-\parts
p^+ p^-) \ ,
 \nonumber \\
 \pb{\pi^\omega,G_\lambda}&=&\frac{1}{2\omega}
 \left(\frac{p^+p^-}{\sqrt{\omega}}-\sqrt{\omega}\right)=
 \frac{1}{2\omega}G_\lambda-\frac{1}{\sqrt{\omega}}
 \ , \nonumber \\
\pb{\pi^\omega,\tmH_\tau}&=&
\frac{1}{2\omega^{3/2}}
(p_-\parts\phi^--p_+\parts\phi^+)-\frac{1}{2\sqrt{\omega}}=
\frac{1}{2\omega}\tmH_\tau-\frac{1}{\sqrt{\omega}}
\ . \nonumber \\
\end{eqnarray}
Now we are ready to analyze the time
evolution of primary and secondary
constraints. Note that the total
Hamiltonian takes the form
\begin{eqnarray}
H_T
&=&\bT_\tau(n_\tau)+\bT_\sigma(n^\sigma)+
\int d\sigma
(\Gamma_\tau\pi^\tau+\Gamma^\sigma
\pi_\sigma + \Gamma_\omega\pi^\omega+
\Gamma_\lambda \pi^\lambda+
\Sigma^\lambda
G_\lambda+\Sigma^\omega\lambda) \ .
\nonumber \\
\end{eqnarray}
Clearly the time evolution of the
primary constraints
$\pi^\tau,\pi_\sigma,\pi^\omega$ is
preserved during the time evolution of
the system. The time evolution of the
constraint $\pi^\lambda$ takes the form
\begin{equation}
\partt\pi^\lambda=\pb{\pi^\lambda,H_T}\approx
-\Sigma^\omega=0
\end{equation}
and hence determines the value of the
Lagrange multiplier $\Sigma^\omega=0$.
In the same way the requirement of the
preservation of the  constraint
$\lambda\approx 0$ during the time
evolution of the system  determines the
value of the constraint
$\Gamma_\lambda=0$. In other words
$\pi^\lambda\approx 0,\lambda\approx 0
$ are the second class constraints.

Now we proceed to the analysis of the
requirement of the preservation of the
constraints
$\bT_\tau(N^\tau),\bT_\sigma(N^\sigma)$
and $G_\lambda$ during the time
evolution of the system.  In case of
$\bT_\tau(N^\tau)$ we find
\begin{eqnarray}\label{bTtau}
\partt \bT_\tau(N^\tau)=
\pb{\bT_\tau(N^\tau),H_T}\approx \int
d\sigma\left[
\frac{N^\tau\Sigma^\lambda}{\omega}\left(
\parts p^-p^+-\parts p^+p^-\right)
+N^\tau
\frac{1}{\sqrt{\omega}}\Gamma_\omega
\right]=0 \ .
 \nonumber \\
\end{eqnarray}
On the other hand the preservation of
the constraint $G_\lambda$ implies
\begin{eqnarray}\label{partGlambda}
\partt
G_\lambda=\pb{G_\lambda,H_T}\approx
-\frac{n_\tau}{\omega} \left(
\parts p^-p^+-\parts p^+p^-\right)-
\Gamma_\omega \frac{1}{\sqrt{\omega}}=0
\ .
\nonumber \\
\end{eqnarray}
Finally the requirement of the
preservation of the constraint
$\pi^\omega\approx 0$ gives
\begin{eqnarray}
\partt \pi^\omega=\pb{\pi^\omega,H_T}
\approx -\int d\sigma
\frac{1}{\sqrt{\omega}}(n_\tau+\Sigma^\lambda)=0
\nonumber \\
\end{eqnarray}
which implies that
$\Sigma^\lambda=-n_\tau$.  Then we
however see that the equations
(\ref{bTtau}) and (\ref{partGlambda})
are not independent. Then one of them
determines the Lagrange multiplier
$\Gamma_\omega$ as
\begin{equation}
\Gamma_\omega=-\frac{n_\tau}{\sqrt{\omega}}
(\parts p^-p^+-\parts p^+ p^-)
\end{equation}
while $n_\tau$ is still free parameter.
 In other words $G_\lambda$
and $\pi^\omega$ are the second class
constraints. The constraint
$\pi^\omega=0$ vanishes strongly while
the constraint $G_\lambda=0$ implies
\begin{equation}
\omega=-p^+p^- \ .
\end{equation}
Finally   we should replace the Poisson
brackets with corresponding Dirac
brackets. In case of the constraints
$\tmH_\tau$ we find
\begin{eqnarray}
\pb{\tmH_\tau(\sigma),\tmH_\tau(\sigma')}_D&=&
\pb{\tmH_\tau(\sigma),\tmH_\tau(\sigma')}-
\nonumber \\
&-& \int d\sigma_1 d\sigma_2
\pb{\tmH_\tau
(\sigma),\pi^\omega(\sigma_1)}\triangle^{-1}(\sigma_1,
\sigma_2)\pb{G_\lambda(\sigma_2),\tmH_\tau(\sigma')}
+\nonumber \\
&+& \int d\sigma_1 d\sigma_2
\pb{\tmH_\tau(\sigma),G_\lambda(\sigma_1)}
\triangle^{-1}(\sigma_1,\sigma_2)
\pb{\pi_\omega(\sigma_2),\tmH_\tau(\sigma')}=
\nonumber \\
&=&\pb{\tmH_\tau(\sigma),\tmH_\tau(\sigma')}
\ ,
\nonumber \\
\end{eqnarray}
where we defined
\begin{equation}
\pb{\pi^\omega(\sigma),G(\sigma')}\approx
-\frac{1}{\sqrt{\omega}(\sigma)}\delta(\sigma-\sigma')
\ , \quad
\triangle^{-1}=-\sqrt{\omega}(\sigma)\delta(\sigma-\sigma')
\
\end{equation}
and we used the fact that all Poisson
brackets are proportional to the delta
functions so that the contributions from
the Poisson brackets between
$\tmH_\tau$ and the second class
constraints vanish.
 In case of the
constraints $\bT_\sigma$ we find that
the Dirac brackets are the same as the
corresponding  Poisson brackets due to
the fact that the Poisson brackets
between $\bT_\sigma$ and the second
class constraints vanish on the
constraint surface. Finally, the Dirac
brackets between $\phi^\pm$ and $p_\pm$
coincide with Poisson brackets due to
the fact that the Poisson brackets
between $\phi^\pm$ and $\pi^\omega$
vanish.

In summary, the reduced phase space is
spanned by the variables
$\phi_\pm,p^\pm$ and $n_\tau,n^\sigma$
together with $\pi^\tau,\pi_\sigma$.
The dynamics is governed by the
Hamiltonian
\begin{equation}
H_{red}= \int
d\sigma(n_\tau\tmH^{red}_\tau+n^\sigma
\mH_\sigma+\Gamma_\tau \pi^\tau+
\Gamma_\sigma \pi^\sigma) \ ,
\end{equation}
where
\begin{equation}
\tmH^{red}_\tau=\frac{1}{\sqrt{-p_+p_-}}
( p_-\parts \phi^--p_+\parts
\phi^+)+2\sqrt{-p_+p_-} \ .
\end{equation}
Note that
$\tmH^{red}_\tau,\mH_\sigma,\pi^\tau,\pi^\sigma$
are the first class constraints and
that the symplectic structure of given
theory is canonical. Finally we should
stress that there are no physical
degrees of freedom  left due to the fact that four
first class constraints listen above
eliminate all physical degrees of
freedom
$\phi^\pm,\pi_\pm,n_\tau,\pi^\tau,n^\sigma,\pi_\sigma$.
%
Explicitly, in order to  fix the
constraints $\tmH_\tau,\mH_\sigma$ we
consider following gauge fixing
functions
\begin{equation}
G^+\equiv \phi^+-k\sigma-\tau\approx 0
\ , \quad G^-\equiv
\phi^--l\sigma-\tau\approx 0 \ ,
\end{equation}
where $k,l$ are integers.  The
diffeomorphism constraint gives
\begin{equation}
p_+k+p_-l=0 \ , \quad
p_+=-\frac{l}{k}p_-
\end{equation}
while the Hamiltonian constraint
$\tmH_\tau=0$ implies
\begin{equation}
p_-=-k \ , \quad  p_+=-l \ .
\end{equation}
Finally note  that the requirement of
the preservation of the gauge fixing
functions $G^\pm\approx 0$  during the
time evolution of the system determines
the value of the variables
$n_\tau,n^\sigma$ which in turns could
serve as the gauge fixing conditions
for the primary constraints
$\pi^\tau\approx 0, \pi_\sigma\approx
0$. Then however the requirement of the
preservation of these gauge fixing
functions determine the value of the
Lagrange multipliers $\Gamma_\tau,
\Gamma_\sigma$. In other words in the
process of the gauge fixing we
completely determined all gauge
symmetry parameters and also all
canonical variables.

\begin{appendix}
\section{Hamiltonian Analysis of $1+1$
dimensional  Scalar Field
Theory}\label{app} In this appendix we
briefly review the Hamiltonian analysis
of $1+1$ dimensional scalar theory
minimally coupled to gravity. In other
words we consider the Lagrangian
density
\begin{equation}\label{lacAp}
\mL_{scal}=-\frac{1}{2}\sqrt{-\gamma}
\gamma^{\alpha\beta}\partial_\alpha
\phi^A\partial_\beta\phi^B \eta_{AB} \
,
\end{equation}
where $A,B=1,2$. Using $1+1$ formalism
we find the Lagrangian density in the
form
\begin{equation}
\mL=\frac{1}{2}\sqrt{\omega}n_\tau
\left(\nabla_n\phi^A\nabla_n\phi^B-\frac{1}{\omega}\parts\phi^A
\parts\phi^B\right)\eta_{AB} \ .
\end{equation}
Then it is easy to find corresponding
Hamiltonian density in the form
\begin{equation}
\mH=\frac{n_\tau}{\sqrt{\omega}}
\mH_\tau+n^\sigma \mH_\sigma+
\Gamma_\tau\pi^\tau+
\Gamma^\sigma\pi_\sigma+\Gamma_\omega\pi^\omega
\ ,
\end{equation}
where
\begin{eqnarray}
\mH_\tau&=&\frac{1}{2}\pi_A
\eta^{AB}\pi_B+ \frac{1}{2}\parts
\phi^A\eta_{AB}\parts \phi^B \ ,
\nonumber \\
\mH_\sigma&=& p_A\parts\phi^A \ .
\nonumber \\
\end{eqnarray}
Observe that now $\omega$ appears in
the combination with $n_\tau$. As a
result the requirement of the
preservation of the primary constraint
$\pi^\omega$ does not generate any
additional constraint. More precisely,
the requirement of the preservation of
the constraints $\pi^\tau\approx
0,\pi^\sigma\approx 0$ implies the
secondary constraints
\begin{equation}
\mH_\tau\approx 0 \ , \quad
\mH_\sigma\approx 0 \ .
\end{equation}
Then the standard analysis shows that
these constraints are the first class
constraints. The gauge fixing of these
constraints eliminate $2$ scalar fields
and corresponding conjugate momenta.

We see that the main difference with
respect to the analysis presented in
the main body of the paper is that in
case of $1+1$ scalar field theory
Lagrangian density (\ref{lacAp})  the requirement of the
preservation of the primary constraint
$\pi^\omega$ does not generate new
additional constraint and hence corresponding
structure of constraints is much simpler.

\end{appendix}

 \noindent {\bf
Acknowledgements:}
 This work   was
supported by the Czech Ministry of
Education under Contract No. MSM
0021622409. \vskip 5mm

\end{document}